\documentclass{article}

\usepackage[a4paper, portrait, margin=1in]{geometry}

\usepackage[authoryear]{natbib}
\usepackage{amsmath}
\usepackage{cases}

\usepackage{graphicx}
\graphicspath{{./images}}

\title{Discounting under inequality and lobbyists disagreement}

\author{Mahdi Mousavi, Mahdi Kohan Sefidi}
\date{Feb. 2023}

\begin{document}
\maketitle
\begin{flushleft}

\end{flushleft}
\begin{abstract}
The appropriate discount rate for evaluating policies is a critical consideration in economic decision-making. This paper presents a new model for calculating the derived discount rate for a society that includes different groups with varying desirable discount rates. The model takes into account equality in society and is designed to be used by social planners. The derived discount rate is a useful tool for examining the social planner's approach to policies related to the future of society. If the discount rate is determined correctly, it can help determine the amount of growth and equality in society, as well as the level of attention paid to long-term public projects.

The model can be customized for different distributions of wealth and discount rates, allowing researchers to extract desired results. Analysis of the model shows that when equality in society is considered, the derived discount rate is lower than the result obtained using Hamilton's method. Social planners must consider that this may increase disagreement in more consuming groups of society at first.
\end{abstract}
\section{Introduction}
One of the challenges policymakers face in the economy for the country's economic development is often when deciding on it, they face challenges in determining the discount rate for long-term investments or long-term public projects. This challenge is such that it must be done in such a way that, while taking into account the level of attention to equality in society and the lack of concentration of capital in the hands of a particular group, development also proceeds in the right direction. One question is whether an index can be created that accurately measures the level of opposition in these groups? When an appropriate discount rate can be extracted, then dissatisfaction levels of groups can be examined through it.

There is substantial evidence in the empirical literature, as indicated by \cite{Feldstein1964}, regarding heterogeneous time preferences. According to \cite{arrow1999}, the ethical concern arises with discounting for projects that yield returns in the distant future (as also mentioned by Ramsey in 1928). However, the question remains as to how to control the level of disagreement among individuals and what tools can be utilized to manage it effectively. The objective is to strike a balance between an acceptable level of disagreement and achieving the desired discount rate for long-term projects.
If we consider individuals' choices regarding discount rates as an ethical matter, as argued by (\cite{Koopmans1960}, \cite{dasgupta2005}, \cite{dasgupta2012}), the question arises of whether these choices are desirable for future generations.

There is considerable debate among economists over the appropriate discount rate for long-term projects, such as climate change mitigation, as highlighted by \cite{stern2007}. Despite the democratic nature of society, the determination of discount rates is subjective and susceptible to bias. These biases are demonstrated by social planners in three out of four cases, with only one out of four cases demonstrating time consistency in their decisions, as noted in a study by \cite{Matthew2015}. Therefore, these biases will influence decision-making in both short and long-term projects. Given these biases, the question arises of whether it is feasible to make ethical and rational decisions that are also in the interest of future generations. This has significant implications for resource allocation and the welfare of future generations.

One method of modeling the discount rate is to use an equivalent discount rate based on different discount rates for different groups within society. According to a study by \cite{Gollier2005}, a method for equalizing discount rates for heterogeneous agents with non-storable income sources was developed. They found that the discount rate "Representative Agent" is a weighted sum of individual rates, where weights depend on the timing of consumption allocation. Gollier also demonstrates that when representatives exhibit a tolerance for consumption volatility (or equivalently, reduce absolute risk aversion), the discount rate RA decreases uniformly over time. However, obtaining an equivalent discount rate assuming exogenous income may not be comprehensive enough for the model, so \cite{Millner2013} proposed a model called "Policy Equivalent" which manages a common pool resource over time through a within-group approach. This model is more suitable for cases that require collective public choices on policies.

In cases where a society needs to make decisions about intertemporal consumption programs for some public goods, a committee is formed at each time period, responsible for determining the consumption level in the current period. Members of each committee have different views regarding the social time preference rate (PRSTP) or the discount rate for municipal services that must be applied to this problem. Some advocates prefer a high discount rate, while others believe that different time periods should be treated equally, and therefore prefer a low discount rate. This issue has been examined by \cite{Millner2018} by considering the Nash equilibrium point in the game with the condition of maximizing growth and consumption for two groups of committee members.

The problem we are examining in this article differs from the usual problem and assumes that the committee responsible for determining the time preference rate must also pay attention to another group in addition to their inherent impact coefficient (such as capital) for wealth redistribution. These groups can be considered as the impact of lobbyists in decision-making, as represented by political representatives.
Furthermore, the social planner aims to reduce inequality in society. As such, they must take into account their development policies while also considering inequality.
The utility model is structured from the Savage framework, which can be examined in \cite{Savage1948}, and has also been used in wealth redistribution by lobbyists by \cite{Zheng2014}. Assuming the level of attention that the representative group makes the decision for, the disagreement created among different groups for the choice is also evaluated, and the optimal cumulative consumption for maximizing the overall utility of society is considered, assuming the existence of each group and the level of attention to inequality by the social planner.

The article will proceed as follows: In section 2, the general steps of model design are presented, which includes calculating the optimal consumption for two, three, and n groups in society, as well as calculating the discount rate for optimal consumption and the level of attention to equality. In section 3, a parametric analysis will be performed, which includes the use of different distributions such as Pareto and gamma distributions for wealth and different discount rates, which are analyzed and discussed. In last sections, the level of disagreement for welfare equivalent and policy equivalent will be examined and compared. In the final section, conclusions will be drawn from the analysis and model presented.

\section{Model}
In this section, we consider a model for the social planner who is responsible for determining the discount rate, taking into account the influence of lobby groups and attention to equality across different groups in society. The first part deals with selecting a model for the utility welfare function that can provide the necessary assumptions for modeling. In the second and third parts, the model's impact is calculated for societies with two and three influential groups, respectively, and their potential applications are explained. Finally, in the last part, the problem is solved with the assumption of $n$ groups in society.

\subsection{Social planner preferences}
One of the assumptions in our modeling is that the social planner intends to regulate the optimal level of consumption in society through the discount rate while reducing inequality. They also aim to measure the disagreement of lobby groups. Due to the nature of the social planner's role in optimizing the allocation of resources across groups, even with the lobbying coefficients, they must still pay attention to inequality. 

Our aim was to develop a model that captures wealth inequality across different groups. Each lobby group, based on their level of influence, as determined by their wealth parameter, can affect the discount rate. The social planner, on the other hand, seeks to optimize the distribution of public resources, and therefore the model must seek an optimal distribution, given this assumption. \cite{Zheng2014}'s model addresses these features in our desired model of social welfare, utilizing a Rawlsian social welfare function. This function seeks to increase the welfare of the most disadvantaged group, requiring special attention to this group. Thus, the model presented in this article emphasizes that in addition to the impact of a particular group's wealth and influence on decision-making, the social planner must also focus on the most disadvantaged group in order to reduce inequality. This special attention is represented by a coefficient.

Assuming that $(x_1, x_2, ..., x_n)$ represents consumption, where $x_i$ denotes the consumption of the $i^{\text{th}}$ interest group. The social planner's preferences over the distribution of consumption among the $n$ interest groups are represented using a Choquet integral, without taking into consideration the internal structure and size of each group. In the general, $n$ groups of lobby we have:
\begin{equation}
W(x_1,x_2,x_3,...,x_n)=(1-\theta)\left(\sum_{i}{y_iU(x_i)}\right)+\theta\min{\{U(x_1),U(x_2),U(x_3),...,U(x_n)\}}
\label{socialplannerModel}
\end{equation}
In utility function $U(.)$ considered as concave function which means $U''(x)<0$, and $y_i>0$, $\sum{y_i}=1$, and $0 \leq \theta \leq 1$.
The well-known Rawlsian egalitarian social welfare function is expressed as:
\begin{equation}
W=\min{\{U(x_1), U(x_2), ..., U(x_n)\}}
\end{equation}
where the parameter $\theta$ represents the social planner's weight on the egalitarian consideration among the $n$ interest groups. Egalitarianism is often viewed as an extreme form of aversion to inequality, and it can also be seen as an attitude towards the inequality of utilities between different interest groups. In the definition of the desirability function of the time discount rate, it is defined as an exponential function and the relationship will be as follows:
\begin{equation}
    U(x_i)=u(x_i)e^{-\rho_i t}
\end{equation}
where $u(x_i)$ consider as utility function that exponential discounted. defined $\rho_i$ as discount rate of each lobbing groups at society, and $0 \leq \rho \leq 1$, $t>0$.

As \cite{Zheng2014} argue, the statement discusses a decision-making framework called the Savage type preference representation (see \cite{Savage1948}), which is used to represent a decision-maker's preferences over different options. The framework considers the decision-maker's subjective belief in probability and attitude towards risk when making decisions under uncertainty.

For equivalent discount rate \cite{Gollier2005} present formulas for calculating these welfare equivalent measures. They demonstrate that when the $x_i$ values are optimally allocated, we observe that these measures hold true. It's important to note that welfare equivalent preferences are determined without any understanding of the factors that generate the group's consumption. Instead, $x$ is treated as a predetermined income flow. Therefore, although welfare equivalent preferences capture the actual value of consumption, they don't account for the impact of preference diversity on the group's collective consumption choices that are determined endogenously. This has implications for economic analysis.

\begin{equation}\label{discount_lobbing}
    \rho^*=\frac{\displaystyle \sum_{i=1}^{n}{\rho_i x_i}}{\displaystyle \sum_{i=1}^{n}{x_i}}
\end{equation}
\subsection{Optimum consumption with two lobbying groups}
Two unequal wealth lobby groups are considered $\{y_1, y_2$\}, with one having a higher wealth coefficient than the other. Assuming these two groups and a utility function \{$U_1, U_2$\} for obtaining optimal consumption, we proceed to solve the equation using the Lagrange method and obtain the maximum consumption value for each group.
\begin{equation}
W(x_1,x_2)=(1-\theta)(y_1U(x_1)+y_2U(x_2))+\theta\min{\{U(x_1),U(x_2)\}}
\end{equation}
as a constrain we have:
\begin{equation}
x_1+x_2=w
\label{eq:x1x2w}
\end{equation}
Lagrange equation for maximization equal:
\begin{equation}
\mathcal{L}=W(x_1,x_2)+\lambda (w-x_1-x_2)
\end{equation}

\begin{equation}
\frac{\partial{\mathcal{L}}}{\partial{x_1}}=\frac{\partial{W}}{\partial{x_1}}-\lambda=0
\end{equation}

\begin{equation}
\frac{\partial{\mathcal{L}}}{\partial{x_2}}=\frac{\partial{W}}{\partial{x_2}}-\lambda=0
\end{equation}

\begin{equation}
\frac{\partial{\mathcal{L}}}{\partial{\lambda}}=w-x_1-x_2=0
\end{equation}

\begin{equation}
\frac{u'(x_1)}{u'(x_2)}=\frac{(1-\theta)y_2}{((1-\theta)y_1+\theta)}
\end{equation}
defined $r$ as $1/\sigma$ and $y_1<y_2$ and $\rho_1>\rho_2$ 
\begin{equation}
x_1=\frac{w((1-\theta)y_1+\theta)^r }{((1-\theta)y_1+\theta)^r+((1-\theta)y_2)^r e^{r(\rho_1-\rho_2)t}} \label{consumption_two_groups}
\end{equation}
To calculate $x_2$, we can extract other parameters of $x$ by substituting the equation in (\ref{eq:x1x2w}). In the final section, the calculated discount rate estimates the discount rate of two lobby groups. Using this discount rate, we can evaluate decision-making in societies where powerful lobby groups in society are divided into two categories. Lobbies have an impact on the consumption structure of governments and this impact can be a tendency to reduce or increase the discount rate (\citealp{Helpman1996, Helpman1992}; \cite{Whinston1986}).

optimum consumption path will be used at calculation equivalent discounting rate like:

\begin{equation}
\rho^*=\frac{\rho_1 x_1 + \rho_2 x_2}{x_1 + x_2}
\end{equation}

In other words, in this hypothetical society, the social planner intends to reduce the rate of inequality while using the coefficient $\theta$ to also control the dissatisfaction of another group and optimize their discount rate (optimization for carrying out public projects for future generations that will have long-term returns).

\subsection{Optimum consumption with three lobbying groups}
In this section, we intend to perform calculations for a society with three lobby groups. In most articles, the binary state of lobbies is examined, but in this section we will examine the ternary state and in the next section we will calculate the optimal consumption value for $n$ lobby groups.Three unequal wealth lobby groups are considered $\{y_1, y_2, y_3$\}, with one having a higher wealth coefficient than the other. Assuming these Three groups and a utility function $\{U_1, U_2, U_3$\} for obtaining optimal consumption, we proceed to solve the equation using the Lagrange method and obtain the maximum consumption value for each group.
\begin{equation}
W(x_1,x_2,x_3)=(1-\theta)(y_1U(x_1)+y_2U(x_2)+y_3U(x_3))+\theta\min{\{U(x_1),U(x_2),U(x_3)\}}
\end{equation}
as a constrain we have:
\begin{equation}
x_1+x_2+x_3=w
\end{equation}
Lagrange equation for maximization equal:
\begin{equation}
\mathcal{L}=W(x_1,x_2,x_3)+\lambda (w-x_1-x_2-x_3)
\end{equation}

\begin{equation}
\frac{\partial{\mathcal{L}}}{\partial{x_1}}=\frac{\partial{W}}{\partial{x_1}}-\lambda=0
\end{equation}

\begin{equation}
\frac{\partial{\mathcal{L}}}{\partial{x_2}}=\frac{\partial{W}}{\partial{x_2}}-\lambda=0
\end{equation}

\begin{equation}
\frac{\partial{\mathcal{L}}}{\partial{x_3}}=\frac{\partial{W}}{\partial{x_3}}-\lambda=0
\end{equation}

\begin{equation}
\frac{\partial{\mathcal{L}}}{\partial{\lambda}}=w-x_1-x_2-x_3=0
\end{equation}

\begin{equation}
\frac{u'(x_1)}{u'(x_2)}=\frac{(1-\theta)y_2}{(1-\theta)y_1+\theta}
\end{equation}

\begin{equation}
\frac{u'(x_2)}{u'(x_3)}=\frac{y_3}{y_2}
\end{equation}

\begin{equation}
\frac{u'(x_1)}{u'(x_3)}=\frac{(1-\theta)y_3}{(1-\theta)y_1+\theta}
\end{equation}

defined $r$ as $1/\sigma$ and $y_1<y_2<y_3$ and $\rho_1>\rho_2>\rho_3$

\begin{equation}
x_1=\frac{w((1-\theta)y_1+\theta)^r}{((1-\theta)y_1+\theta)^r+((1-\theta)y_2)^r e^{r(\rho_1-\rho_2)t} + ((1-\theta)y_3)^r e^{r(\rho_1-\rho_3)t}}
\end{equation}

optimum consumption path will be used at calculation equivalent discounting rate like:

\begin{equation}
\rho^*=\frac{\rho_1 x_1 + \rho_2 x_2 + \rho_3 x_3}{x_1 + x_2 + x_3}
\end{equation}
One of the applications of this model is that in elections, parties (lobby groups) may have different approaches from the people and therefore we have divided the lobby groups into two lobbies and one majority group of people. This equation can be used to calculate the equivalent utility and also the level of disagreement created for different discount rates. These discount rates are important in public projects where a specific group has an effect on decision-making.

\subsection{Optimum consumption with $n$ lobbying groups}
In this section, we examined a larger number of lobby groups or committees. We used the Lagrange multi-variable optimization method to determine the optimal consumption if we increase the number of groups to $n$. Our goal was to create a summarized relationship that could optimally examine $n$ groups and calculate the discount rate. This relationship can help decision-making groups improve their performance and consider different income distributions for members and each person's desired discount rate.

Other articles have tried to find a desirable point between committees resulting from a game with Nash equilibrium like (\cite{Millner2018}). However, in this article, we simulated the existing space without considering the game between groups. Our aim is to see the impact of the decisions made by the social planner group on the society's discount rate and compare it with other discount rates of constituent groups.
\begin{equation}
W(x_1,x_2,...,x_n)=(1-\theta)(\sum_{i=1}^{n} y_iU(x_i))+\theta\min{\{U(x_i)\}}
\end{equation}
Also if we consider $y_1<y_2<...<y_n$, and $\rho_1>\rho_2>...>\rho_n$, utility function is considered as an common utility function $U(c)=c^{(1-\sigma)}/1-\sigma$.
Optimum consumption path is solved in below. To facilitate solving the equation, we define the variable as $\alpha_i$ as follows:
\begin{equation}
\alpha_i=\frac{(1-\theta)y_1+\theta}{(1-\theta)y_i}  \text{ where $i \neq 1$}
\label{eq:alphai}
\end{equation}
By solving the optimization equation for the variable $x_1$ and its relationship with other $x_i$ parameters, we arrive at the following relationship:
\begin{equation}
\frac{x_i}{x_1}=\left(\frac{e^{\rho_1-\rho_i}}{\alpha_i}\right)^{\frac{1}{\sigma}} \text{ where $i \neq 1$}
\label{eq:xix1}
\end{equation}
After solving the equations for all consumption values for each member under the condition that $i \neq 1$, we arrive at the following relationship which will greatly help us in further solving:
\begin{equation}
\frac{x_n}{x_m}=\left(\frac{y_n/y_m}{e^{\rho_n-\rho_m}}\right)^{\frac{1}{\sigma}} \text{ where $n \neq 1$, $m \neq 1$ and $n \neq m$}
\label{eq:xnxm}
\end{equation}
with equations (\ref{eq:alphai}), (\ref{eq:xix1}) and (\ref{eq:xnxm}) we solved optimization equation for $x_1$ and $x_i$ as below:

\begin{align}
x_1&=\frac{w((1-\theta)y_1+\theta)^r }{\displaystyle ((1-\theta)y_1+\theta)^r+\sum^{n}_{i=2}((1-\theta)y_i)^r e^{r(\rho_{1}-\rho_i)t}} \label{eq:full_x1}\\
x_i&=\frac{w}{\displaystyle 1+\left(\frac{\alpha_i}{e^{(\rho_1-\rho_i)t}}\right)^r+\sum_{j>1, j \neq i}^{n}\left(\frac{y_j/y_i}{e^{(\rho_j-\rho_i)t}}\right)^r} \text{ where $i \neq 1$} \label{eq:full_xi}
\end{align}
If we assume extreme conditions for the time variable, the calculated optimal consumption values are for $t = 0$ and $t \rightarrow \infty$. For $t \rightarrow \infty$, the answer is $x_1 = 0$. For $t = 0$, the answer is:
\begin{equation}
x_1=\frac{w((1-\theta)y_1+\theta)^r }{((1-\theta)y_1+\theta)^r+\sum^{n}_{i=2}((1-\theta)y_i)^r}
\end{equation}
For other groups, when $t \rightarrow \infty$ then $x_i = w$, when $t = 0$ we have:
\begin{equation}
x_i=\frac{w}{\displaystyle 1+\alpha_i^r+\sum_{j>1, j \neq i}^{n}\left(y_j/y_i\right)^r} \text{ where $i \neq 1$}
\end{equation}
The optimum consumption path will be utilized in the calculation of the equivalent discounting rate as follows:
\begin{equation}
    \rho^*=\frac{\displaystyle \sum_{i=1}^{n}{\rho_i x_i}}{\displaystyle \sum_{i=1}^{n}{x_i}}
\label{eq:rhostar}
\end{equation}
The equivalent utility equation using the derived discount rate is equal to:
\begin{equation}
    U^*(x)=u(x)e^{-\rho^*t}
    \label{eq:uitilityEqui}
\end{equation}
The equation calculated in relation (\ref{eq:rhostar}) is a derivative of the discount rate for $n$ groups of society. To obtain the optimal utility value, it is sufficient to calculate the utility function exponentially in the value of the derived discount rate according to relation (\ref{eq:uitilityEqui}). Millner's article presents a relationship in relation (\ref{eq:utilityMillnerEqu}) that differs significantly from the equation calculated here. Millner's approach assumes no effect of lobbying and advances only in the general process with Nash equilibrium under an assumption that differs from ours. Millner considers zero as an extreme value for a short time interval of maximizing groups' problem in part of Hamilton's optimization equation. The relationship calculated by Millner can be evaluated under different conditions compared to our approach and may serve as a suitable basis for future research.

The consumption choices made by committees that aggregate the preferences of their members in a utilitarian manner are described. Specifically, it is assumed that at time $t$, a committee adopts a social choice rule represented by a function $W_t$, where:
\begin{equation}
W_t=\sum_{i}{y_iV_{it}}
\end{equation}
Heterogeneous opinions on the appropriate value of the PRSTP are held by committee members, i.e. indices $i$, $j$ exist such that $\delta_i \neq \delta_j$. $\delta_i$ is interpreted as $i$'s normative opinion on the appropriate rate of social impatience.
If it exists, the limit equilibrium consumption path is observationally equivalent to the optimal path:
\begin{equation}
  \hat{ \delta } = r + \eta (A-r)
\end{equation}
where $\eta>0$ is the elasticity of marginal utility. Equilibrium welfare is believed by member $i$ at time $\tau$ to be
\begin{equation}
V_{i\tau} = 
\begin{cases} 
\frac{(S_{\tau}A)^{1-\eta}}{1-\eta}\frac{1}{(r-A)(\eta-1)+\delta_i} & \text{if } \eta \neq 0 \\
\frac{1}{\delta_i}\ln{(S_\tau A)}+\frac{1}{\delta_i^2}(r-A)       & \text{if } \eta = 0
\end{cases}
\label{eq:utilityMillnerEqu}
\end{equation}

\section{Parametric analysis}
In this section, to analyze the behavior of the equivalent discount rate function, we examine the equation by determining some variables such as wealth distribution and discount rate distribution among individuals in society. We assign a distribution for each and then examine the behavior of the equivalent discount rate. Next, we calculate the value of the disagreement parameter for the social planner problem. In another section, assuming that the social planner wants to examine their action with respect to the growth rate using an equivalent policy structure, we calculate and compare the amount of disagreement and analyze trends based on changes in the rate of influence of the lobby group.

\subsection{Pareto distribution of wealth}
Pareto distributions are valuable tools for modeling and predicting in many socioeconomic contexts. However, there is a distinct advantage in concentrating on one specific area of application: the distribution of wealth sizes. This is the context in which Vilfredo Pareto used them (\cite{Barry2015}, \cite{Pareto1964}). I want consider distribution of wealth ($y_i$) as Pareto distribution, total form of paretor distribution function is like:
\begin{equation}
f(x) =
 \alpha  k^{\alpha } x^{-\alpha -1} \quad \text{where $x\geq k$}
\end{equation}
Normalized of $f(i)$ is
\begin{equation}\label{eq:normal}
y_i=\frac{f(i)}{\displaystyle \sum_{i=1}^{n} f(i) }
\end{equation}
Consumption with Pareto distribution at $t=0$:

\begin{align}
x_1&=\frac{w((1-\theta)(y_1)+\theta )^r }{((1-\theta)(y_1)+\theta)^r+\sum^{n}_{j=2}\left((1-\theta)(y_j)\right)^r}\label{eq:x1pareto}
\\
\\
x_i&=\frac{w}{\displaystyle 1+\alpha_i^r+\sum_{j>1, j \neq i}^{n}\left(y_j/y_i\right)^r} \text{ where $i \neq 1$}\label{eq:xipareto}
\end{align}
Given that various operations are performed on the Pareto distribution parameter, denoted by $y$ in equations (\ref{eq:x1pareto}), (\ref{eq:xipareto}), it is not easy to calculate its analytical and explicit response. Therefore, we used the Monte Carlo numerical estimation method for this purpose (\cite{monteCarloMethods}).
\begin{figure}[h]
    \centering
    \begin{minipage}{0.45\textwidth}
        \centering
        \includegraphics[width=\textwidth]{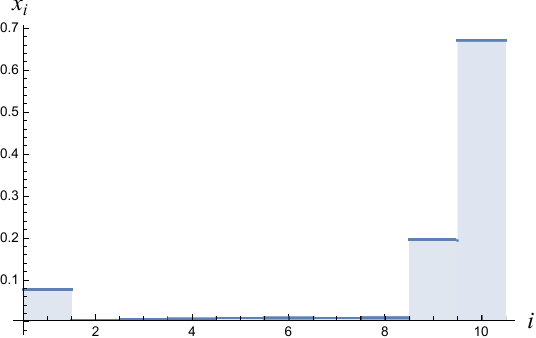}
        \caption{Distribution of consumption ($x_i$) of each groups}
        \label{fig:image1}
    \end{minipage}\hfill
    \begin{minipage}{0.45\textwidth}
        \centering
        \includegraphics[width=\textwidth]{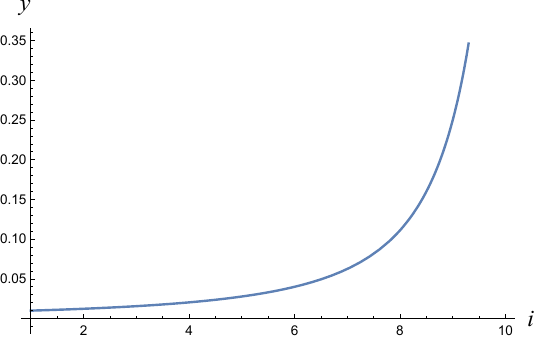}
        \caption{Pareto distribution of wealth ($y_i$) of each groups as $1/x^2$ where sorted Ascending for plotting per i}
        \label{fig:image2}
    \end{minipage}
\end{figure}
Figure \ref{fig:image1} shows the results of calculating consumption values for the given distribution and input data. As can be seen in the chart, considering the impact coefficient $\theta$, the consumption value of group $x_1$ does not follow the general trend and has increased more than the first 8 groups. This increase indicates the impact of the social planner group's decision on equality in society. The increase in consumption of the first group is significant with respect to the decision-making group. Figure \ref{fig:image2} shows the wealth distribution for all groups in order from 1 to 10. The wealth distribution function in this analysis is considered as $1/x^2$. The $y$ values are also normalized in the chart and the relationship is used.
\begin{figure}[h]
    \centering
    \includegraphics[width=0.65\textwidth]{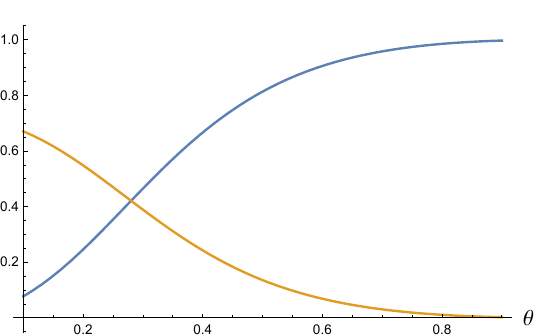}
    \caption{Changing value of consumption (x) by $\theta$}
    \label{fig:image3}
\end{figure}
A change in $\theta$ will affect the consumption of each group shown in Figure \ref{fig:image3}. The figure displays the consumption of the influential lobbying group influenced by $\theta$ and the consumption of other groups (which exhibit similar behavior overall). An increase in $\theta$ results in an increase in consumption for the influential lobbying group and a decrease for other groups. The orange graph represents the consumption of the influential group and the blue graph represents the consumption of one of the other groups.

One point that this graph can show us is the effectiveness of creating an impact factor for a group to reduce inequality and how it can significantly increase the consumption of that group in society. Another point that may be better analyzed in the next section is that the Social planner should not support a group for a long time by determining the value of $\theta$, as this would cause a large difference between this group and other groups in society and have a negative impact on equality. It can also create dissatisfaction in society and push society towards monopolization.

\subsection{Gamma distribution of discounting}
In the preceding section, we examined the utility of lobbying groups solely in terms of their consumption. Consumption is related to the wealth of a group, represented by $y_i$, with the group possessing the least wealth having the minimum utility. The distribution of wealth among groups follows a Pareto distribution.

In this section, we investigate the impact of discount rates on the consumption behavior of lobbying groups. Discount rates are represented by $\rho_i$ and their distribution among groups follows a Gamma distribution, as described in \cite{gamma2001}. Additionally, we explore how changes in the equality coefficient, represented by $\theta$, affect the consumption behavior of an influential group. Given that wealth distribution and discount rate distributions are not independent random variables and our intention to use Monte Carlo simulation (\cite{monteCarloMethods}), we assume a constant distribution for wealth among groups and consider the distribution of discount rates as a Gamma distribution.

\begin{equation}
\displaystyle
f(x) =
 \frac{\beta ^{-\alpha } x^{\alpha -1} e^{-x/\beta}}{\Gamma (\alpha )} \quad \text{where $x > 0$}
\end{equation}
In order to calculate consumption, equations (\ref{eq:full_x1}) and (\ref{eq:full_xi}) were simulated using random discount variables. When $y_i$ is assumed to have the same value for all groups, with $\sum_i y_i = 1$ condition, the equation (\ref{eq:full_xi}) is modified accordingly:
\begin{equation}\label{eq:xi_same_yi}
x_i=\frac{w}{\displaystyle 1+\left(\frac{\alpha_i}{e^{(\rho_1-\rho_i)t}}\right)^r+\sum_{j>1, j \neq i}^{n}\left(e^{r(\rho_i-\rho_j)t}\right)} \text{ where $i \neq 1$}
\end{equation}

\begin{figure}[h]
    \centering
    \begin{minipage}{0.45\textwidth}
        \centering
        \includegraphics[width=\textwidth]{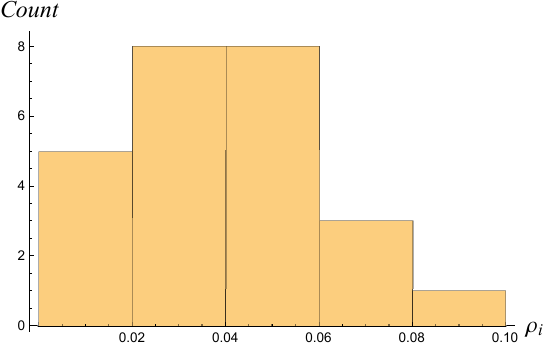}
        \caption{Histogram of discount rate ($\rho_i$) of each groups (25 groups)}
        \label{fig:image4}
    \end{minipage}\hfill
    \begin{minipage}{0.45\textwidth}
        \centering
        \includegraphics[width=\textwidth]
        {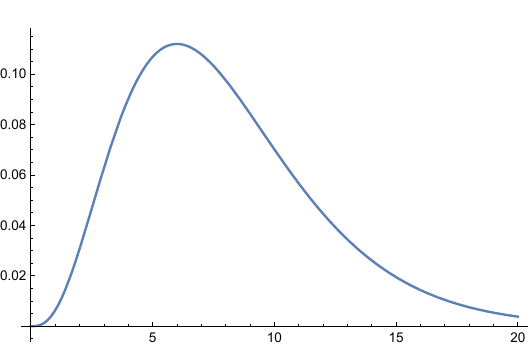}
        \caption{Gamma distribution of discount rate ($\rho_i$) of each groups as $1/96.\exp{(-x/2)}.x^3$}
        \label{fig:image5}
    \end{minipage}
\end{figure}

For normalization, equation (\ref{eq:normal}) was also utilized. According to Figure \ref{fig:image5}, the parameters selected for the distribution are $\alpha=4$ and $\beta=2$, resulting in the distribution equation $f(x)=(1/96).x^3.\exp{(-x/2)}$. A total of 25 random values were used on the distribution and the histogram of the results is shown in Figure \ref{fig:image4}.

\begin{figure}[h]
    \centering
    \begin{minipage}{0.45\textwidth}
        \centering
        \includegraphics[width=\textwidth]{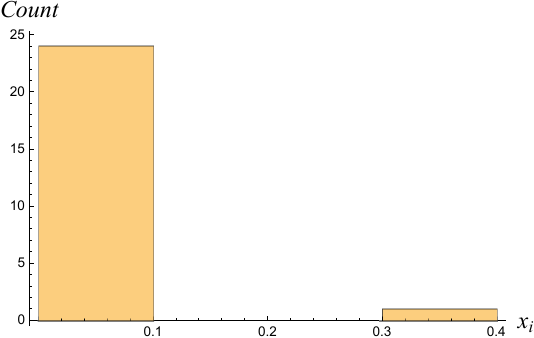}
        \caption{Histogram of consumption ($x_i$) of each groups}
        \label{fig:image6}
    \end{minipage}\hfill
    \begin{minipage}{0.45\textwidth}
        \centering
        \includegraphics[width=\textwidth]{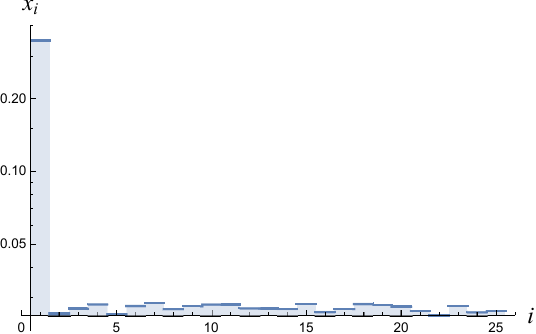}
        \caption{Consumption of each groups (result of Mont Carlo simulation)}
        \label{fig:image7}
    \end{minipage}
\end{figure}

In Figure \ref{fig:image5}, the discount rate distribution is given for a gamma distribution function. The distribution function equation is considered as $1/96.\exp(-x/2).x^3$. This distribution function is randomly created for 20 groups and can be seen in Figure 4. Now, assuming this distribution function for influential groups and using the Monte Carlo method, we can see the resulting consumption rate for different groups of society in Figures \ref{fig:image6} and \ref{fig:image7}. Figure \ref{fig:image6} shows the consumption distribution between groups and this distribution fully explains the difference created between different groups as a result of the social planner's attention.
Figure \ref{fig:image7} shows the consumption rate of each group after placing the distribution function in the equation and you can see how the social planner's attention to group $x_1$ will affect consumption and the significant difference created in this group's consumption rate. Unlike the previous section where an unequal distribution was considered for society, in this section the wealth distribution parameter has become ineffective and, in other words, represents wealth equality among individuals (groups).

\begin{figure}[h]
    \centering
    \includegraphics[width=0.65\textwidth]{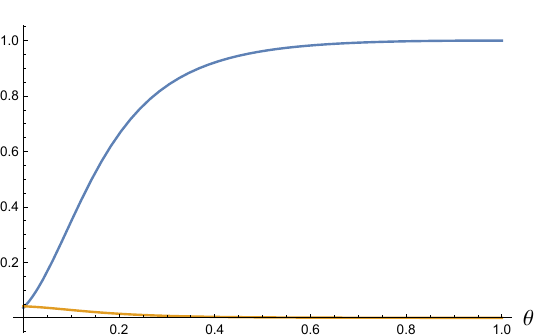}
    \caption{Changing value of consumption (x) by $\theta$}
    \label{fig:image8}
\end{figure}

In \ref{fig:image8}, we observe the changes in consumption as a function of the changes in the impact of $\theta$, without taking into account the level of inequality considered. The aim of this analysis was to purely examine the impact of $\theta$ without the presence of other variables. In this figure, as can be seen, the consumption level of the group that has been considered by the social planner has shown a significant difference and a high slope of consumption increase, which is even greater than the amount set in the previous section. The effectiveness of this tool is also proven in this section, and since the discount rate distribution in society has been proposed as gamma, we can better observe its impact in this analysis.

In such circumstances, it can be seen how much influence has changed in group 1 by the attention that the social planner has paid to it and this change has been very intense. From this section, many results can be shown in political decision-making as well as dynamic conditions relative to decision-making. Changing trends to support equality policies must be fully and controlledly applied to the system to prevent differences and severe imbalances in society. In continuation, similar to the previous section, Figure \ref{fig:image8} examines the change in consumption of group $x_1$ with another group and this difference as well as the sharp growth of group consumption in those early sections of the chart. It should be noted that this chart exaggeratedly shows the effect of lobbying in societies with greater equality in that when equality increases with a little attention to a specific group, dissatisfaction in other groups increases at a higher rate and intensity and thus distances society from monopoly. In some articles about consolidating democratic structures (\cite{Houle2009}), some things have been said that can also be inferred from this chart. The more equal a society is, the more attention to a specific group increases dissatisfaction in other groups at a higher rate and intensity and thus distances society from monopoly.

\subsection{Disagreements based on inequality}
Lobbying is the act of attempting to influence decisions made by government officials, often by private interest groups or individuals. 
The tendency of social planner in decreasing inequality can significantly impact the consumption patterns of groups within political systems. By advocating for policies that favor certain industries or products, this tendency can lead to changes in the availability, price, and demand for certain goods and services. As a result, the consumption patterns of different groups may diverge.

This divergence in consumption patterns can lead to disagreement among groups. For example, if the tendency of social planner in decreasing inequality results in policies that favor one industry over another, the consumption patterns of groups associated with those industries may change. This can lead to disagreement among those groups about the fairness or effectiveness of those policies.

Analyzing the effect of the tendency of social planner in decreasing inequality on consumption and its impact on disagreement could provide valuable insights into the dynamics of political systems and their influence on economic behavior. Understanding how this tendency affects consumption and how this in turn affects disagreement among groups can help policymakers make more informed decisions about how to balance the discount rate of different groups.

When analyzing equation (\ref{socialplannerModel}) without considering the equality parameters, where $\theta=0$, the equation takes on a simplified form. This has been solved by other researchers.
\begin{equation}
    V(x_1,x_2,x_3,...,x_n)=\sum_{i}{y_iU(x_i)}
\end{equation}

To solve this state, we utilize the concept of welfare equivalent. Welfare equivalent preferences are utilized to gauge the welfare of a group at each point in time. The group's aggregate consumption is distributed efficiently and its instantaneous welfare is represented by a function. \cite{Gollier2005} derived expressions for welfare equivalent time preferences and tolerance for consumption fluctuations. Welfare equivalent preferences are defined independently of the process that generates the group's consumption. In this case, we simply treat $X_t$ as an exogenous given income stream. 

A comparison of welfare equivalent under scenarios with and without the equality parameter could provide valuable insights into the impact of inequality on the well-being of different groups.
\begin{align}
V(X_t,t)&=\max_{x_{ti}}{\sum_{i}{y_i U_i(x_{ti})}e^{-\rho_it}} \text{ s.t. } \sum_{i}{x_{ti}=X_t} \label{millner_vx}\\
T^V(X_t,t)&=-\frac{V_{X}}{V_{XX}}=\sum_{i}T_i(x_i) \label{millner_tx}\\
\rho^V(X_t,t)&=-\frac{V_{tX}}{V_{XX}}=\frac{\sum_{i}\rho_i.T_i(x_t)}{\sum_{i}T_i(x_t)} \label{millner_rv}\\
V(X_t,t)&=U(X_t)e^{-\rho^Vt} \label{millner_vxe}
\end{align}

Using a previously defined parameter such as $\theta$ to represent the tendency of social planner in decreasing inequality on consumption, we could compare the welfare equivalent of the entire population under scenarios with $\theta=0$ and $\theta>0$. This comparison could provide valuable insights into how inequality affects the well-being of different groups and how this in turn affects disagreement among them.

\begin{equation}
    U(x)=\frac{x^{(1-\eta)}}{1-\eta} \label{common_felicity_fun}
\end{equation}
\begin{equation}
    T(x)=\frac{x}{\eta}
\end{equation}
Considered $X=1$, shows that in this special case:
\begin{equation}
    x_{it}=y_i e^{-\rho_i t}
\end{equation}
\begin{equation}\label{discount_welfare}
    \rho^V=\frac{\displaystyle \sum_i{\rho_i(y_i e^{-\rho_i t}})}{ \displaystyle \sum_i{(y_i e^{-\rho_i t}})}
\end{equation}
In our study, we considered three lobbying groups that are representative of society. We then assigned values for discount rates and their wealth distribution as $y_i=\{0.5, 0.3, 0.2\}$ and $\rho_i=\{0.01, 0.02, 0.03\}$. For other variables, we set $t=1$ and plotted $\theta>0 \text{ to } 1$. By substituting these values into equations (\ref{discount_welfare}) and (\ref{discount_lobbing}), we obtained the results shown in Figure 2.

\begin{figure}[h]
    \centering
    \begin{minipage}{0.45\textwidth}
        \centering
        \includegraphics[width=\textwidth]{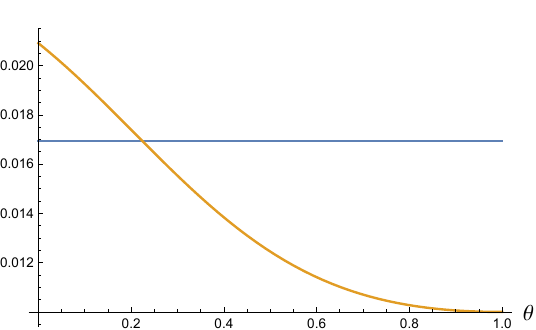}
        \caption{Change of equivalent discount rate by change of $\theta$}
        \label{fig:image9}
    \end{minipage}\hfill
    \begin{minipage}{0.45\textwidth}
        \centering
        \includegraphics[width=\textwidth]{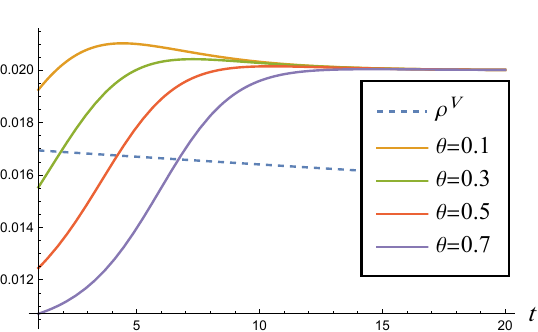}
        \caption{Change of discount rate by unit of time per $\theta$}
        \label{fig:image10}
    \end{minipage}
\end{figure}
In Figure \ref{fig:image9}, the orange graph represents the equivalent discount rate based on changes in $\theta$, while the blue graph shows the calculated equivalent interest rate for these three groups by welfare equivalent. As we can see, the equivalent discount rate calculated in this article is larger when $\theta$ is small and decreases as $\theta$ increases. Interestingly, for these three groups, if we increase $\theta$, the discount rate continuously decreases. An important point to note is that this discount rate is a function of time and we need to see how it changes over time, which Figure \ref{fig:image10} will better illustrate.

When we examine the continuous colored lines in Figure 10, we observe that the discount rate changes over time and its value increases over time and converges after a certain point. As the discount rate increases with respect to $\theta$, the graph appears to shift to the left. This shift causes the discount rate values for lower $\theta$ values to increase and for higher $\theta$ values to decrease. However, an important point here is that the slope of the discount rate graph increases with respect to the importance of creating equality in society. This can cause changes in consumption rates and may create dissatisfaction for some and satisfaction for others. The dashed blue line also shows the changes in the welfare equivalent over time, which decreases over time and follows a linear trend towards decrease.

A general point that can be derived from this graph is that if a social planner wants to allocate a median level of attention to inequality ($\theta=0.2$), it can be observed that consumption in society increases after a short period of time. This increase in consumption even exceeds the welfare equivalent and may cause dissatisfaction in the long run for a group in society that is looking for longer-term growth rates.

Another point to consider is that the decision to determine the discount rate should be dynamically changed after correcting inequality and creating balance between groups. With this approach, we will always see lower discount rates with better equality indices in society. It should be noted that reducing the discount rate to create equality at the beginning of the path may make groups in society with higher consumption tendencies more dissatisfied than groups with lower short-term consumption tendencies. An interesting point is that more deprived groups that should receive more attention may initially become more dissatisfied. This point may cause a decrease in their voting rates even from groups whose interests are supposed to be strengthened. On the other hand, lobby groups also work to prevent this issue from reducing their overall desirability.

\subsection{Lobbing groups and simple growth model policy equivalent discounting}
In our model, we incorporate varying growth tendencies for each group. Our approach to modeling these groups is grounded in the discount rate and neoclassical growth theory for social planners. We used a Cobb-Douglas production function to generate output from the stock of capital (S), while taking into account depreciation. We assume a constant labor supply and no technical progress, meaning that output production is solely affected by the productivity of the capital stock. We compare equivalent discounting rate related to the consumption of each of the lobbying groups to policy equivalent representative agent with simple growth function $F(S)$.
\begin{equation}
    F(S)=S^{\gamma}-\delta S \label{growth_function}
\end{equation}
According to \cite{Millner2013}, policy equivalent preferences are formulated in such a way that the representative agent's optimal consumption plan aligns with the social planner's equilibrium. Our objective is to provide a characterization of these preferences. Let us consider a representative agent (RA) with an instantaneous utility function denoted by $R(X, t)$. We do not impose any restrictions on $R(X, t)$ except that it must be increasing and concave with respect to $X$ and satisfy the condition that $\lim_{X \to 0} \partial R(X, t)/ \partial X$ for all t. The RA will solve this problem.

\begin{equation}
R(X_t, t)=\max_{X_{t}}{\sum_{i}{y_i{U_i(x_{ti})}e^{-\rho_it}}} \text{ s.t. }\dot{S}=F(S)-\sum_{i}x_{ti} \label{millner_vrp}
\end{equation}

Consider groups indexed by the parameter $i$, each with an associated welfare weight $y_i \in (0,1)$ such that $\sum_{i}y_i=1$. The initial value of the resource stock is given by $S(0)=S_0$. These agents exhibit additive time-consistent preferences. The utility derived by individual $i$ from consumption $x_{ti}$ at time $t$ in the future is given by $U_i(x_{ti})e^{-\rho_it}$, where $U_i(.)$ represents the individual's utility function and $\rho_i$ denotes their pure rate of time preference. We assume that their income is derived from a resource $S$, with a growth function given by $F(S)$. According to \cite{Millner2013}, the representative agent's consumption equation can be solved for the optimal allocation using equations (\ref{millner_rv}) and (\ref{millner_vrp} to \ref{millner_jr}).
\begin{align}
    \rho^R(X_t,t)&=(1-J^R(X_t))F'(S_t)+J^R(X_t,t)\rho^V(X_t,t) \label{millner_ror} \\
    J^R(X_t,t)&=\frac{\sum_i {T_i(x_{ti})}}{T(\sum_i x_{ti})} \label{millner_jr} 
\end{align}
In our analysis, we utilized the common felicity function as indicated in Equation (\ref{common_felicity_fun}). Additionally, we employed the optimum consumption function of two lobbying groups as shown in Equation (\ref{consumption_two_groups}) then values of the agents' consumption:
\begin{equation}
    x_{ti}=\left(\frac{e^{-\rho_i.t}.y_i}{\lambda}\right)^{1/\eta}
\end{equation}
Calculating each parameters then substituting (\ref{millner_jr}) into (\ref{millner_ror}) shows that in this special case:
\begin{align}
J^R(X_t, t)&=\frac{x_{t1}^2+x_{t2}^2}{(x_{t1}+x_{t2})^2}=\frac{\rho_1(e^{-\rho_1t}y_1)^{2/\eta}+\rho_2(e^{-\rho_2t}y_2)^{2/\eta}}{((e^{-\rho_1t}y_1)^{1/\eta}+(e^{-\rho_2t}y_2)^{1/\eta})^{2}} \\
\rho^V(X_t, t)&=\frac{\rho_1x_{t1}^2+\rho_2x_{t2}^2}{x_{t1}^2+x_{t2}^2}=\frac{\rho_1(e^{-\rho_1t}y_1)^{2/\eta}+\rho_2(e^{-\rho_2t}y_2)^{2/\eta}}{(e^{-\rho_1t}y_1)^{2/\eta}+(e^{-\rho_2t}y_2)^{2/\eta}}\\
F'(S)&=\zeta S_{t}^{\zeta-1}-\delta
\end{align}
The functions $F'(S_t)$, $\rho^V(c_t,t)$, and $T^V(c_t,t)$ are all exogenous given.
\begin{figure}[h]
    \centering
    \includegraphics[width=0.65\textwidth]{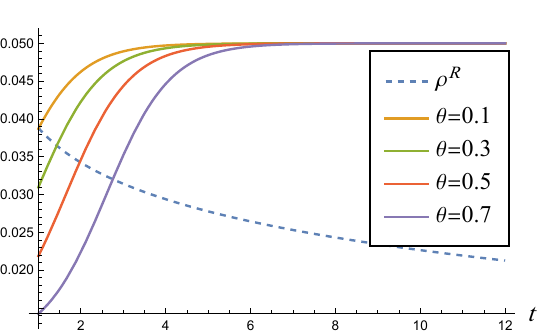}
    \caption{Changing discount rate by $t$ for equivalent discounting by $\theta$ and policy equivalent}
    \label{fig:image11}
\end{figure}
When growth is considered in the Policy equivalent model compared to the welfare equivalent, the discount rate values in time 1 will be higher and therefore the rate of decrease will also be high along the way. As can be seen in the graph, the calculated discount rates for different $\theta$s are also higher than before, but all have considered lower numbers for growth. In other words, this graph states that if we want to pay attention to equality in society and at the same time have growth, we must consider lower discount rates than Policy equivalent but overall higher than welfare equivalent.

Another important point in the graph is the slope of the discount rate graphs relative to different $\theta$s. The slope of these graphs is also higher than the previous case, so the rate of consumption change grows sharply and therefore we must consider the previous policy of considering the amount of attention to equality dynamically to prevent sudden growth in the discount rate.

The overall model of improving equality alongside growth states that the considered discount rate must be large numbers and therefore this large discount rate may not be desirable for a group of society who prefer low discount rates and may cause dissatisfaction for them. Also, if these policies continue continuously and without review for specific groups, it may cause problems such as a decrease in growth rate and an increase in dissatisfaction among other strata of society and an increase in the Gini coefficient after some time after improvements.

\section{Conclusion}
In this article, an attempt was made to present a new model for calculating the derived discount rate for a society with the assumption of a society that includes different groups with different desirable discount rates and also pays attention to equality in society by the social planner. The derived discount rate is one of the suitable ideas for a general but simplified examination that can show the social planner's approach to policies related to the future of a society. If the discount rate for society is determined correctly, it can determine the amount of growth and equality in society as well as the amount of attention to long-term public projects (for future generations).

One of the features that the model can provide due to its calculation for n groups in society is the ability to change different distributions of wealth and discount rates, so researchers can use it for customized calculations and extract the desired results.

In democratic societies, due to the ethical assumption of human rights in determining consumption and investment for the future, balance must be considered in this field and therefore under this logic, the presented model can be helpful to some extent with all simplifications. Also, when it is necessary to measure a policy before implementation, a decision is made using this model to examine the amount of disagreement created for different groups in society.

The points extracted from the analysis of the model also showed that the amount of discount rate when paying attention to equality in society is less than the total result of the social planner problem optimized by Hamilton's method in various researches and therefore this point must be considered by the social planner that it may increase disagreement in the more consuming group of society at first. On the other hand, continuing in reducing equality policy without applying dynamic discounting will increase inequality and the derived discount rate will increase sharply.


\medskip
\bibliography{main}
    
\end{document}